# Electrorheological suspensions of laponite in oil: rheometry studies under steady shear

*K. P. S. Parmar, Y. Méheust, and J. O. Fossum[*]*

Department of Physics, Norwegian University of Science and Technology, NTNU,

7491 Trondheim, Norway

[*]jon.fossum@ntnu.no

**Abstract:** We have studied the effect of an external DC electric field (~kV/mm) on the rheological properties of colloidal suspensions consisting of aggregates of laponite particles in a silicone oil. Microscopy observations show that under application of an electric field greater than a triggering electric field $E_c$ ~ 0.6 kV/mm, laponite aggregates assemble into chain- and/or column-like structures in the oil. Without an applied electric field, the steady state shear behavior of such suspensions is Newtonian-like. Under application of an electric field larger than $E_c$, it changes dramatically as a result of the changes in the microstructure: a significant yield stress is measured, and under continuous shear the fluid is shear-thinning. The rheological properties, in particular the dynamic and static shear stress, were studied as a function of particle volume fraction, for various strengths(including null) of the applied electric field. The flow curves under continuous shearing can be scaled with respect to both particle fraction and electric field strength, onto a master curve. This scaling is consistent with simple scaling arguments. The shape of the master curve accounts for the system's complexity; it approaches a standard Herschel-Bulkley model at high Manson numbers. Both dynamic and static yield stress are observed to depend on the particle fraction $\Phi$ and electric field $E$ as $\Phi^\beta E^\alpha$, with $\alpha$ ~1.85, and $\beta$~1 and 1.70, for the dynamic and static yield stresses, respectively. The measured yield stress behavior may be explained in terms of standard conduction models for electrorheological systems. Interesting prospects include using such systems for self-guided assembly of clay nano-particles.

# 1. Introduction

Electrorheological (ER) fluids [1-9] are colloidal suspensions consisting of high dielectric constant- and/or high conductivity- particles in an insulating liquid. Rheological properties (i.e. viscosity, shear stress, viscoelastic modulii) of such suspensions can easily be controlled by application of an external electric field of about 1.kV/mm. The electric field induces a polarization of each particle, and interactions [10] between polarized particle aggregates lead them to form chains and /or- columns like- structures [11-12] in the suspensions. This structuring phenomenon only observed in the presence of an external electric field is termed the ER effect. Such changes in rheological properties are know to be reversible and rapid with characteristic time scales typically of the order of 1-100 ms for a field strength of about 1.kV/mm. Under steady state shear, such structuring produces very high shear stresses in the suspensions. At rest, the micro-structure jams the flow, resulting in a high yield stress that must be exceeded for the fluid to flow. The quick response to the applied field and the resulting high yield stress has attracted much interest, triggering many scientific and industrial research studies. Many devices have been proposed based on ER effect: clutches, brakes, damping devises, hydraulic valves etc. [2-3, 7, 18, 19, 17].

Dielectric constant, particle conductivity [8-9], volume fraction of particles [5, 8-9, 13], as well as the nature of the applied electric field (frequency and magnitude) [5, 8-9], are all parameters found to be major factors controlling the shear stress behavior of ER fluids. Other factors such as particle geometry [14] (i.e., shape [15-16] and size [17-18]), and polydispersity [17-18], are also considered critically important in some types of ER fluids. Sometimes a small amount of additives like water [19-20] ad- sorbed and/or ab- sorbed on the particles also plays a major role on the shear stress behavior of an ER fluid. The water-activated ER fluids will generally malfunction at high- or low- operating temperature, so it is always desirable to (i) use suspensions whose ER mechanism is not based on ad-/ab-sorbed water, and (ii) to remove the water content of the ER fluid.

Without electric fields, the ideal steady state shear behavior of an ER fluid is Newtonian-like, i.e there is no yield stress [21, 22] and the apparent viscosity (ratio of shear stress to shear rate) does not depend on the shear rate, but only on the volume fraction, $\Phi$, of the particles. In particular, the $\Phi$ - dependent viscosity of colloidal suspensions of monodisperse spherical particles in a Newtonian-liquid of viscosity $\eta_0$ are often approximated, for $\Phi<3\%$, by the Batchelor [23] or Krieger-Dougherty [23] relations, which are $\eta = \eta_0 \left(1-\Phi/\Phi_{max}\right)^{-2.5\,\Phi_{max}}$ and $\eta = \eta_0 \left(1+2.5\,\Phi+6.2\,\Phi^2\right)$, respectively. These relations account for the interactions between the particles themselves and between particles and the surrounding liquid. One empirical equation which has been found to account for the viscosity of monodisperse and polydisperse sphere suspensions [25], in a range of particle fraction up to 50% is:

$$\eta = \eta_0 \left(1+\frac{0.75}{\frac{\Phi_m}{\Phi}-1}\right)^2 , \quad (1)$$

in which $\Phi_m$ corresponds to the particle fraction at maximum packing. The three relations above can in general be used to describe the viscosity of moderately-concentrated colloidal suspensions of isotropic, monodisperse, and non-charged particles [23-25], although relation (1) has been found to be an appropriate description for concentrated suspensions of kaolinite clay[45], whose particles are anisotropic and polydisperse. Under application of a sufficiently large electric field ($E > E_c \sim$ 1kV/mm), ER fluids show well-defined yield stresses, beyond which they tend to be shear-thinning (pseudoplastic) i.e. they exhibit a viscosity that decreases with increasing shear rate. At moderate shear rates at which the shear-thinning behavior can be discarded, the typical steady-shear behavior of an ER fluid under these conditions is most often characterized as a Bingham-like solid [21] given by the expression

$$\tau = \tau(E)+\eta_{pl}\,\dot{\gamma} \text{ for } \tau > \tau(E)$$
$$\dot{\gamma} = 0 \text{ for } \tau < \tau(E) \quad (2)$$

where $E > E_c$ is the intensity of the applied electric field, $\tau(E)$ is the dynamic yield stress at field strength $E$, $\dot{\gamma}$ is the shear rate, and $\eta_{pl}$ is the $E$-dependent plastic viscosity, which approaches the suspensions viscosity at sufficiently high shear rate. Equation (2) suggests that ER fluids exhibit a solid-like behavior below their dynamic yield stress and a liquid-like behavior above it. Above the dynamic yield stress, and at larger shear rates, the Bingham behavior is often replaced by a Herschel-Bulkley model that accounts for the shear-thinning behavior. It should also be noted that an absolute yield stress is in itself an elusive property, as there are many ways to evaluate the yield stress for a fluid-like substance, and no single best technique can be identified [22]. One common method of measuring a yield stress is to extrapolate the shear stress versus shear rate curve back to the shear stress intercept at zero shear rate. The value obtained by this method is termed the dynamic yield stress. It can be strongly influenced by the data-range of shear rates used, and by the rheological model selected to do the extrapolation. The static yield stress, on the other hand, is the shear stress needed to initiate shear flow of a fluid that is initially at rest [25].

Under application of a large electric field, the dynamic yield stress measured for an ER fluid under continuous shear can be significantly different from its static yield stress. This situation is illustrated in Fig. 1, in which we show, as a red continuous line, the changes in the flow curve of an ER fluid that is forced to flow at an increasing shear rate, from a configuration at which it is at rest.

Experimentally, and theoretically, it is found that the magnitude of the static yield stress is critically dependent on the volume fraction of particles, $\Phi$, and on the magnitude of the applied electric fields, $E$. The yield stress of an ER fluid is generally characterized by the relation

$$\tau_0(\Phi, E) \propto \Phi^\beta E^\alpha \qquad . \qquad (3)$$

Exponent values $1 \leq \alpha \leq 2$ are commonly observed. The polarization theory [12, 27-28], based on the mismatch in dielectric constant between the particles and the suspending liquid, predicts the exponents to be $\alpha = 2$ and $\beta = 1$. The conduction theory and non-linear conduction theories [29-35], on the other hand, are based on the mismatch in conductivity between the particles and the suspending liquid. They predict an exponent α in the range 1≤α< 2, and: (i) β<1 at both low- and high- volume fractions of particles, but (ii) β>1 at intermediate volume fractions of particles [9, 21]. In all these models, the polarization and conductivities are assumed to depend linearly on the strength of the applied electric field. The models can be refined by taking into account many-body effects, i.e., by introducing multipolar interactions that better describe the interactions at relatively high particulate volume fractions [25, 35].

In this paper, we study the ER behavior of suspensions consisting of laponite particle aggregates suspended in a silicone oil. We study their rheology under electric fields larger than about 2 $E_c$, using controlled shear rate (CSR) tests, under continuous shearing, and controlled shear stress (CSS) tests on static suspensions. The experimental method is described in section 2, while the results are presented and discussed in section 3. Concluding remarks are given in Section 4.

## 2. Experiments

**Materials:** We prepared suspensions of Laponite RD clay particles, purchased in powder form from Laporte Ltd. (UK). In a dilute aqueous suspension, the individual laponite clay particle is a disc of thickness 1 nm and of average diameter 30 nm [37-38]. These discs have a negative charge on their surface, due to the substitution of low charge ion in their crystal structure, and they have a small positive charge on their edges due to unsatisfied broken bonds. The surface charge density of individual discs is 0.4 e- /unit cell [37-38], the specific particle density is 2.65 [39].

Due to that negative structural surface charge, primary particles of laponite in the dry state are stacks of several discs with charge-balancing interlayer $Na^+$ ions between discs. In a dry laponite powder, primary particles form aggregates whose size can vary from about a few nm to a few μm. The size of such aggregates can be further reduced by finely grinding the laponite powder either through a milling procedure, or manually with the help of a mortar and spatula. In the present case, the laponite powder was finely grinded by the latter method.

A silicone oil (a Newtonian liquid) Dow Corning 200/100 Fluid (dielectric constant of 2.5, viscosity of 100 mPa.s and specific density of 0.973 at 25 $^O$C) was used as a suspending liquid because it is relatively non-polar and non-conductive (compared to laponite), with a DC conductivity of the order of magnitude of $10^{-12}$ S/m [34-35].

**Sample Preparation:** Seven suspensions of laponite RD in silicone oil were prepared with different volume fractions, in the following steps: (1) the appropriate amounts of laponite powder and silicone oil were chosen; (2) traces of any water or moisture contents of both the laponite powder and silicone oil were eliminated by heating at 130 $^0$C for 72 hr; (3) heated laponite powder and silicone oil were immediately mixed in glass tubes which were then sealed; (4) the sealed glass tubes were left to cool down to room temperature; (5) each glass tube was vigorously shaken by hands for ~5 min; (6) the glass tubes were placed in an ultrasonic bath for 30 min at 25 $^O$C and again vigorously shaken by hands; (7) after step 6, aggregates of laponite particles were well-dispersed in silicone oil without any sediment or floc large enough to be distinguishable to the eye; optical microscopy observations showed that the aggregates were three-dimensional, retaining little trace of the original platelet geometry of the individual laponite particles, and that the size distribution for their largest dimension ranged between 0.5 and 2 micrometers.

**Microscopy Observations:** Visual observations were carried out using a special purpose microscopy sample cell consisting of two parallel and identical 1/2 mm-thick copper electrodes separated by a gap of 2 mm, and glued onto a standard transparent quartz glass microscopic slide[40]. The gap between the electrodes is closed at its ends by a non-conducting plastic material. The top part of the cell is open, and the sample cell was mounted horizontally on a stereomicroscope connected to a digital camera and to a PC. Changes in the structure of laponite suspension (<1ml) placed between copper electrodes were recorded using MS Windows-compatible software.

**Rheological Measurements:** The rheology of our laponite suspensions was measured under DC electric fields using a Physica MCR 300 Rotational Rheometer equipped with a coaxial cylindrical cell Physica CC27/ERD, specially designed for ER measurements. The cell has an outer cylinder diameter of 14.46 mm and an inner cylinder diameter of 13.33 mm. The immersion length of the inner cylinder is 40 mm, and the corresponding sample volume is 19.35 ml. Two grounding brushes connected to the internal cylinder's axis induce an artificial ~1 Pa yield stress in all data, but this value is negligible compared to all yield stress values addressed here. All rheological measurements were carried out at constant temperature (25 $^0$C). Two types of rheology tests were performed:

- Steady shear rheological properties were measured after the suspensions had been pre-sheared at $\dot{\gamma}$ =200 s$^{-1}$ for 60 s, in order to impose the same initial conditioning to the samples before each measurement run. These tests are termed "Controlled Shear Rate (CSR) tests".
- To determine the static yield stress, Controlled Shear Stress (CSS) Tests, in which a linearly increasing shear stress (in steps of 2 Pa) was imposed, were conducted in disrupt suspensions after they had been subjected to the external electric field for 300 s.

**3. Results and discussions**

**3.1 Microscopy observations**

Microscopy pictures of a laponite suspension of concentration Φ = 17.9 % (v/v), placed between two copper electrodes, are shown in Fig. 2. For $E$ ~ 0.0 kV/mm, the laponite aggregates are randomly dispersed in the silicone oil (Fig. 2 (a)). As the magnitude of the applied electric field approaches a triggering limit $E_c$ of approximately 0.6 kV/mm, a slow motion of particle aggregation is observed in the suspension, and laponite aggregates start to assemble into chain- and/or column-like structures parallel to the direction of applied electric field. With increasing magnitude of applied electric field ($E$), the polarization of the laponite particles increases, causing faster aggregation into chain or- column like structures.

For the lowest particle volume fraction ($\Phi \approx 18\%$) of prepared laponite suspensions, the complete capture of laponite particles into ER structures to form a fully stable static ER fluid occurs in a few tenths of seconds (t < 40 sec), at ambient temperature and under an applied electric field $E = 2.0$ kV/mm (see Fig. 2(b)).

### 3.2 CSR tests ─ Steady-state shearing

The flow curves for different volume fractions of laponite suspensions at zero applied electric field are shown in Fig. 3(a). All suspensions exhibit a Newtonian-like behavior, i.e. a constant apparent viscosity (i.e., ratio of shear stress to shear rate) and no distinguishable preyield stress. The viscosity of the suspensions increases with the volume fraction $\Phi$ of the laponite particles, as illustrated in Fig. 3(b). As $\Phi$ increases, the various interaction forces - hydrodynamics forces, dispersion forces, electrostatic forces and polymeric forces (i.e., short range repulsion forces due to oil molecules possibly adsorbed on the particles) - simultaneously act between the particles themselves, and between particles and the surrounding silicone oil. The net result is an increase in the viscosity. The $\Phi$-dependent viscosity curve (Fig. 3 (b)) can not be approximated to the Batchelor relation [23] or the Krieger-Dougherty relation [23] (see section 1), because of the complexity of our suspensions and their large particle fraction. The relative magnitude of various interactions is affected by the particle properties such as large variation in surface roughness, size and shape polydispersity, and heterogeneous charge distribution [23]. In the present case the properties of the initial laponite aggregates are not well known, so a clear distinction cannot be made between these relative magnitudes of interaction forces that control the rheology of suspensions. The empirical relation (1) does not describe the present zero E-field viscosity data either; however a fit to a similar relation of the form [25]

$$\eta = \eta_0 \left(1 + \frac{K}{\frac{\Phi_m}{\Phi} - 1}\right)^2, \qquad (4)$$

with K= 0.234 and $\Phi_m$=76.5%, fits the data in Fig. 3(b) well. The fitting parameters depend only weakly on the data range used to do the fitting, indicating that relation (4) holds for all investigated particle fractions.

Fig. 4(a) and 4(b) show the flow- and viscosity- curves, respectively, for a laponite suspension of particles volume fraction $\Phi$=35.3 %, under various strengths of the applied electric field. Application of an electric field causes a strong increase in shear stress (and viscosity). Furthermore, the flow curves become pseudoplastic (or shear-thinning) over the entire range of shear rates, as shown by the decrease of viscosity in Fig. 4(b). At low shear rate, the suspensions exhibit a dynamic yield stress that increases with the electric field; this is a behavior commonly observed in ER fluids.

These flow curves can be understood in terms of the dynamics of the formation and rupture of the columnar structures. The dipolar interaction between particles is responsible for the cohesion of the ER structures, which tends to hinder the flow. On the other hand, hydrodynamic forces caused by the applied shear tend to destroy the ER structures in order to promote the flow. At high shear rates and/or at low applied electric field, the effect of hydrodynamic forces (shearing) on the structural reformation process is expected to become much stronger than the electrostatic forces. In this limit, the ER fluid is completely broken down, i.e., no chain-like ER structure remains, and the suspension is expected to behave like a Herschel-Bulkley fluid. This is observed in Fig. 4(a) at large shear rate, for the lowest investigated value of the field. The higher the field strength, the larger the limit shear rate above which this phenomenon occurs. More precisely, the dipole-dipole interaction is proportional to the square $E^2$ of the electric field intensity; that of the shear strength acting on a particle within an ER chain is proportional to the local shear constraint, and hence scales proportionally to the shear rate $\dot{\gamma}$ (assuming that there is a linear velocity field across the gap of the Couette cell). Hence, a normalized shear allowing to compare the relative impact of shear on the cohesive ER structures, between different runs, is $\dot{\gamma}/E^2$, which is proportional to the classic Mason number $M_n$ [8-9]. Thus, we have rescaled the horizontal axis in Fig. 4(a) by $E^2$, and subsequently found visually the scaling factors $s(E)$ to apply on the vertical axis in order to have all flow curves fall on top of each other. In this manner a nice collapse (not shown in Fig. 4) of all data is found. The dependence of $s(E)$ on $E$ is shown in the inset of Fig. 4(c). For the 4 lowest values of the electric field strengths ($2E_c \leq E \leq 5E_c$), this dependence is nicely fitted by a power law with an exponent 1.86. Thus we propose a scaling of the flow curves with respect to the electric field in the form

$$\tau(E) = E^{1.86} \, g\!\left(\frac{\dot{\gamma}}{E^2}\right) , \qquad (5)$$

in which the function $g$ is a master curve accounting for the rheological behavior of the suspensions at large enough shear rate. The flow curves normalized according to Eq.(5) are presented in Fig. 5(c); they collapse nicely with each other, except for those recorded at the two largest values of the electric fields (consistently with the inset). Note that the "peaks" in the flow curves at $\gamma/E^2 \simeq 40$ USI denote the lower shear rate at which the master curve can still be considered a proper common description of the flow curves. Below this value, the collapse of the flow curves is more arguable, probably because in this range of shear rates, the intrinsic complexity of the system, in terms of surface roughness of the laponite aggregates, polydispersity in size and shape of these aggregates, and heterogeneous charge repartition of the particles, play an important role.

In Fig. 5, we show flow curves obtained under an applied electric field strength of $E = 1.8$ kV/mm, for various particle volume fractions $\Phi$. Here we have found scaling factors visually, both along horizontal and the vertical axis, and looked for a simple scaling law accounting for them. If we simply consider that an increase in the particle fraction $\Phi$ results in an average distance between clay particles that scales as $d \sim \Phi^{-1/3}$, then, considering that the electric dipole-dipole interaction scales as $1/d^2$, the cohesion of the chains is expected to scale as $\Phi^{2/3}$, and a scaled shear rate accounting for the influence of the particle fraction would be $\dot\gamma/\Phi^{2/3}$. From our data, the scaling factors found along the horizontal axis are nicely fitted by a power law with an exponent 0.64, quite close to that rough estimate. Besides, the scaling factors along the vertical axis provide a vertical scaling in the simple form $\tau/\Phi$, that holds for particle volume fractions 20%<$\Phi$<40%. We therefore propose a scaling of the flow curves with respect to the particle fraction in the form:

$$\tau(\Phi) = \Phi\ h\left(\frac{\dot\gamma}{\Phi^{0.64}}\right), \qquad (6)$$

where $h$ is a generic master curve. The scaled flow curves are shown in Fig. 5(b). As expected, the collapse is satisfying for 20%<$\Phi$<40%. Above this particle fraction, the vertical scaling is not optimal any more, and it seems that another scaling is reached for $\Phi$ >45%. Note that there can be an optimum volume fraction of particles for good ER fluids, corresponding to a maximum in yield stress value at a given applied electric field [2-6]. Above this optimal particle fraction, self-crowding, for example by several aggregates forming larger aggregates, may restrict yield stress values which then do not increase with increasing magnitude of the applied electric field. Such a behavior is consistent with the scaling of Eq. (6) not holding above a certain particle fraction, although this has not been investigated in detail here.

The scaling relation (6) is observed to hold independently of the strength of the electric field. Taking both scaling relations (5) and (6) into account, we propose a general scaling relation for the flow curve in the form

$$\tau(\Phi,E) = \Phi\ E^{1.86} f\left(\frac{\dot\gamma}{\Phi E^2}\right). \qquad (7)$$

The master curve $f$ contains the rheological behavior for the ER suspensions of particle volume fraction in the range [20%; 40%], subjected to an external electric field in the range [$2E_c$; $5E_c$]. The scaling is only approximate for $\dot\gamma/E^2 < 40$ USI, but by considering that it holds on average at lower shear rates, we can interpret the dynamic yield stress as the limit at zero shear rate of relation (7). We then obtain a dependence of the dynamic yield stress consistent with Eq. (3), with exponent values $\alpha$=1.86 and $\beta$=1.

Note, however, that the conduction model, which provides a similar dependency, deals with static configurations, i.e., it is only suited to determining a static yield stress, which we address in the following section.

### 3.3 CSS tests − Static yield stress determination

As discussed in section 1, when a static suspension, under application of an electric field, is subjected to an increasing and controlled shear stress, the rupture of ER structures occurs above some critical stress known a static yield stress. Below this static yield stress, flow does not occur. For practical purposes, as previously discussed an illustrated in Fig. 1, the true yield stress for an ER fluid is its static yield stress, not the dynamic yield stress addressed in section 3.2. We show in Fig. 6(a) the CSS test flow curves recorded from a suspension of particle fraction 35.8%, under various electric field strengths. The static yield stress values has been extracted from such curves, as the stress value at which a rapid jump in shear rate is observed, and the measured shear rate becomes significant. The static yield stresses increase with the electric field, which is consistent with what we observed for the dynamic yield stress, as presented in section 3.2. Our observed values of static yield stress are comparable to or smaller than (see Table 1) those of other known ER fluids [4, 9, 13, 16], such as mica [16], hematite [41], saponite [42], zeolite [43], and a recently-discovered ER fluid with a giant electrorheological effect [44].

CSS tests as those presented in Fig. 6(a) were carried out with suspensions of various particle fractions. In Fig. 6(b), we have plotted the obtained static yield stresses as a function of the applied electric field, for all the fractions investigated. The plots are well fitted by a power law relation with an exponent $\alpha$ in the range [1.72-1.94], depending on the particular volume fraction $\Phi$. All values obtained for $\alpha$ at the different volume fractions of laponite particles are summarized in Table 2.

As the static yield stress is intrinsically an elusive quantity (see end of section 1), the values measured experimentally can display considerable variations. Therefore we consider that the variations in the exponent values of Table 2, which show no clear trend as a function of the volume fraction and do not even depend monotonically on it, are rather due to the experimental uncertainty. We choose to describe the dependence of the static yield stress as a function of $E$ in terms of a simple power law with an exponent 1.85 that is the average of the exponents given in Table 2. Note that this value of 1.85 is very close to that previously obtained for the dynamic yield stress: 1.86. In Fig. 7, we have plotted the data of Fig. 6(b), normalized by $E^{1.85}$, as a function of the particle fraction. The plots collapse onto each other, and we fit to the overall data a power law with an exponent 1.70. We finally obtain a dependence of the static yield stress in accordance with the conduction model, i.e. to Eq. (3) with exponents α=1.85 and β=1.79. These values of the exponents α suggest that the particles' conductivity increases with increasing electric fields and has to be taken into account together with the pure dielectric constant in order to properly explain the rheological behavior of our laponite-based suspensions. The possible movement of associated/intercalated cation ($Na^+$) and migration of surface charges may cause the induced polarization of these particles. The exponents value β>1 may be attributed to our working regime of moderate to high volume fractions.

## 4. Conclusions

We have studied the electrorheological behavior of suspensions consisting of laponite particles in silicone oil with various particle fractions ranging from ~20 to ~50%. A critical triggering DC electric field $E_c$ ~0.6 kV/mm was observed, sufficient to polarize the laponite particles and have them form chains and/or- columns like-structures. In the absence of an external electric field, the rheology of the suspensions shows Newtonian-like behavior, and changing the particle volume fraction only increases the viscosity. The Φ-dependent viscosity can be well fitted to the empirical relation given by Eq. (4).

We have studied the rheology under application of an electric field $E>2\ E_c$. Under continuous shear, the flow curves approach an ideal Herschel-Bulkley model at large Mason number, and can be rescaled with respect to both the electric field strength and the particle fraction onto a common master curve for $\dot{\gamma}/E^2 > 40$ USI, $2\ E_c<2\ E<5\ E_c$ and 20%<Φ<40%. The overall behavior is shear-thinning, as expected. CCS tests have also been carried out, and have provided a dependence of the static yield stress on the particle fraction and electric field in the form $\tau_0(\Phi, E) \propto \Phi^{1.70} E^{1.85}$, where the scaling exponent for $E$ is the same as that obtained for the scaling of the flow curves under continuous shear. These results suggest that, despite the intrinsic complexity of the system under study, its rheology is not so different from systems of spherical particles addressed by the classic conduction models, in which interaction forces are governed by the dielectric and conductivity mismatch between the dipolar particles (laponite) and the suspending liquid (silicone oil). This reflects the fundamental three-dimensionality of the dispersed laponite particle aggregates, as opposed to the intrinsic two-dimensionality of the individual particles; this three-dimensionality is supported by optical microscopy images. The shape of the master flow- curve under CSR tests (Fig. 4(c) and 5(b)) accounts for the complexity of the suspensions' rheology. In particular, it displays an apparent peak or inflexion point at $\dot{\gamma}/E^2 \sim 30\text{-}40$ USI. We are at present not able to explain this feature, other than regarding it as resulting from a subtle interplay between two or more of the forces at work in this system: dipolar, hydrodynamic, and polymeric.

Future prospects of this work include expanding the rheology studies to lower field strengths than the ones reported here, both below and above the triggering field. In addition, studying the same system by scattering techniques (X-ray, neutron or visible light), may allow us to obtain further information about aggregate shapes and their relative positioning within the ER structures. Furthermore, one very interesting prospect for all future studies would be to attempt functionalizing laponite surfaces in order to control the particle dispersion and thus possibly enable studies of systems of individually-dispersed laponite particles. In view of the numbers cited in Table 1, such laponite-based ER fluids would not be able to compete with other systems in terms of practical use, owing to the moderate value of the yield stress. Rather, in addition to possibly serving as a good physical model system for ER fluids based on anisotropic particles, one application of such systems of functionalized laponite particles could be in guided self-assemblies of nanoparticles for nano-templating and/or inclusion in composite materials.

## 5. Acknowledgements

*The authors acknowledge the help of G. Helgesen for determining the sizes of the laponite aggregates in the silicone oil. This work was supported by NTNU, and by the Research Council of Norway (RCN) through the RCN NANOMAT Program as well through a RCN SUP (Strategic University Program) project granted to the Complex CRT (Collaborative Research Team) in Norway (www.complexphysics.org).*

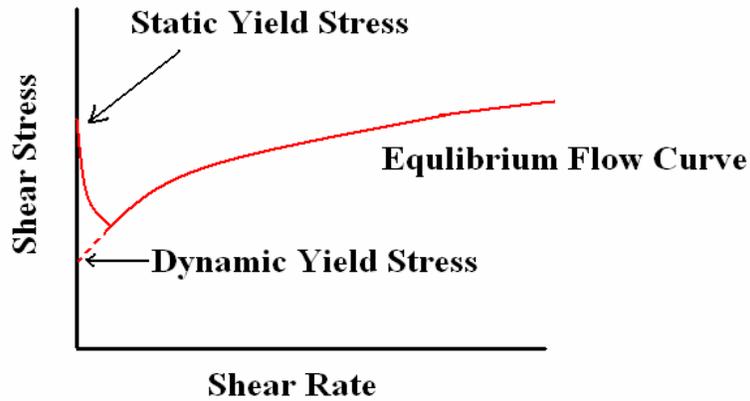

**Fig. 1:** Static and dynamic yield stress. The dynamic yield stress is the yield stress for a completely broken down (i.e., subjected to continuous shearing) ER fluid, whereas the static yield stress is the yield stress for an undisrupted ER fluid. In an ER fluid, both quantities can be very different: the full red curve in this figure shows typical rheology changes when forcing increasing shear rates onto an ER fluid initially at rest.

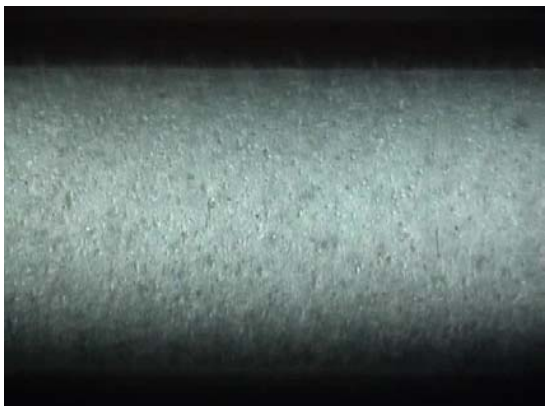
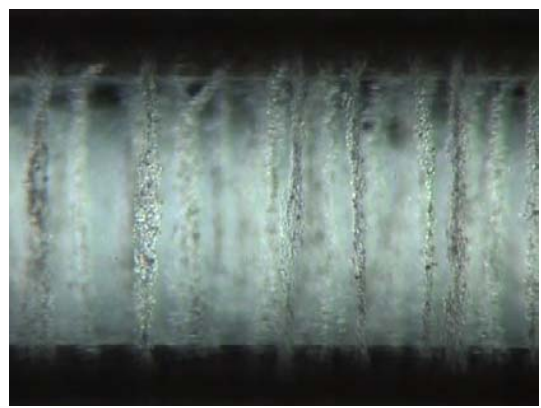

(a) at time, t = 0 s  (b) at time, t ~ 40 s

**Fig. 2:** Effect of an electric field (E) on the particle arrangement in a suspension of laponite particles suspended in silicone oil, of concentration $\Phi \simeq 18\%$. The two black regions are uniformly-spaced (1mm) copper electrodes glued on a transparence microscopic glass slide: (a) Static ER fluid under E ~ 0; clay particles are randomly dispersed. (b) Static ER fluid under E = 2.0 kV/mm; chain- or column-like structures of clay particles have formed.

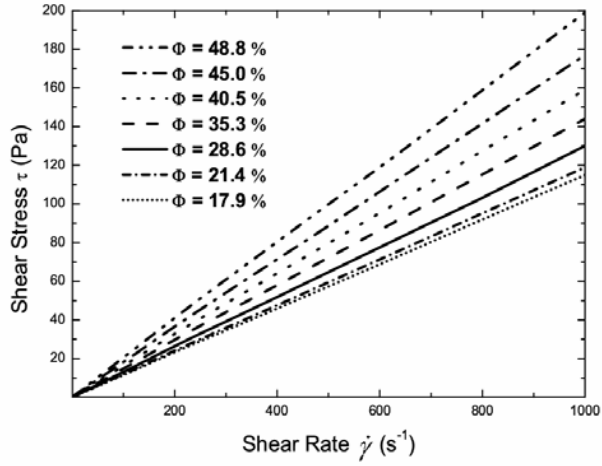 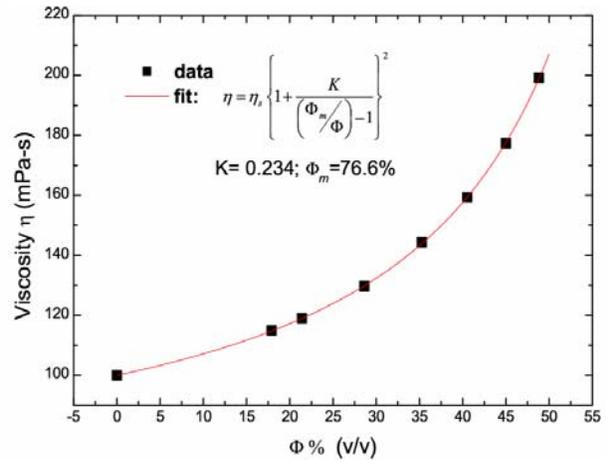

**(a)** **(b)**

**Fig. 3: (a)** Flow curves of laponite suspensions with different volume fractions, with no external electric field applied. **(b)** Viscosity vs. volume fraction curve of laponite particles. The viscosities are the slopes of the straight lines in Fig 3(a). An empirical relation resembling that of Chong et. al.[25]. (see Eq. (4)) has been fitted to the data.

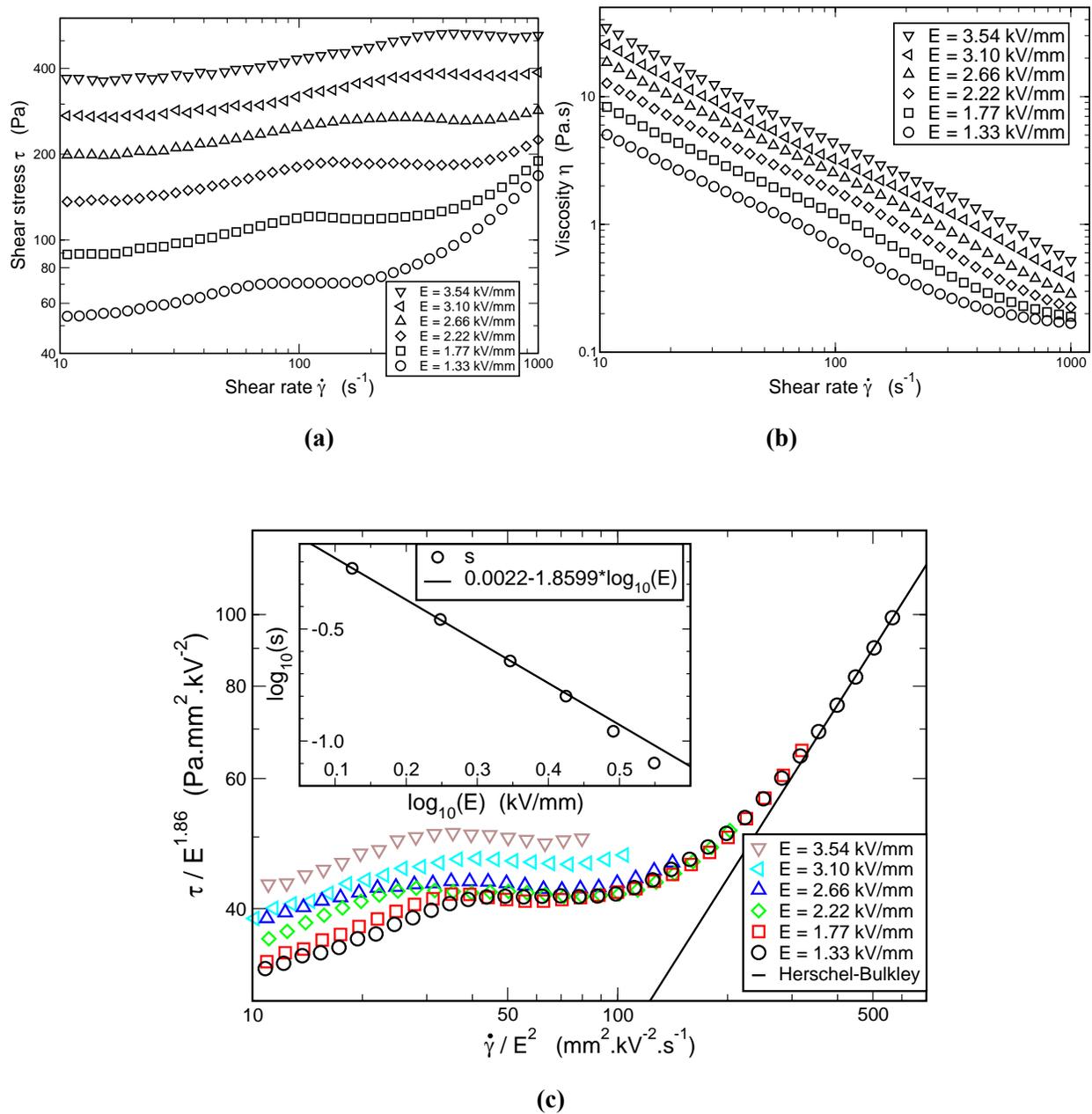

**Fig. 4: (a)** Flow curves (log-log plot) for a suspension of volume fraction Φ=35.3 % under various magnitudes of the applied electric field. **(b)** Corresponding curves of viscosity vs. shear rate. **(c)** Flow curves renormalized according to Eq.(5) (log-log plot); the Herschel-Bulkley model to which the rheological behavior converges at large shear rates is shown as a dashed line. The inset shows the vertical scaling factors found visually, and how they are fitted by a $E^{1.86}$ power law for field strengths $2 E_c < E < 5 E_c$.

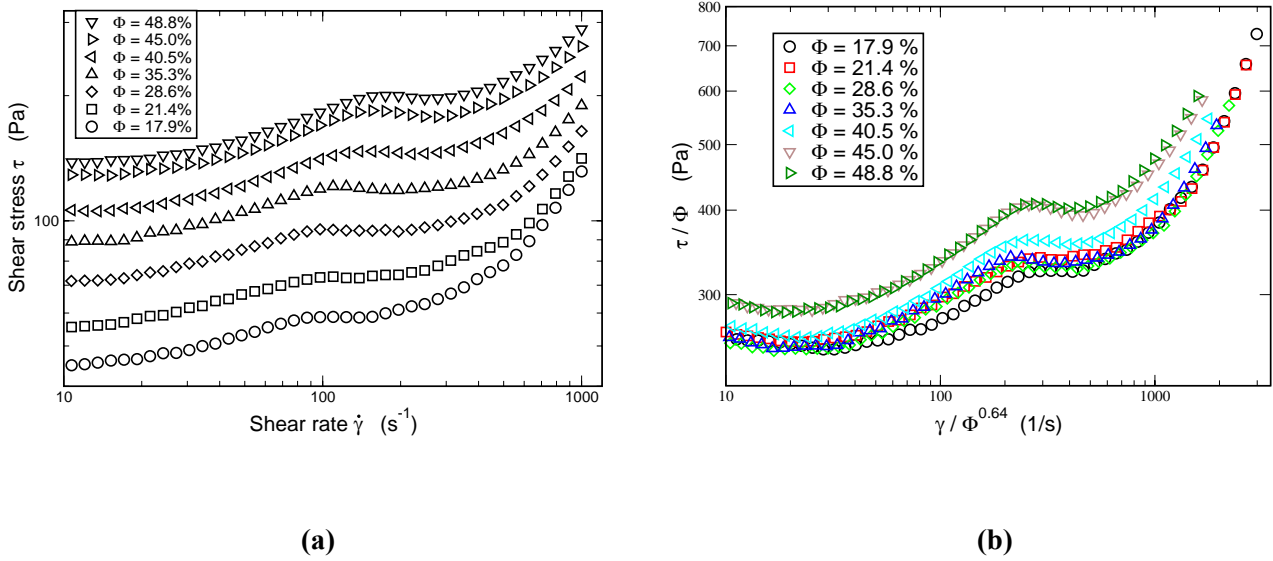

**(a)** **(b)**

**Fig. 5: (a)** Flow curves at a fixed strength, E=1.8 kV/mm, of the applied electric field, for various volume fractions of laponite particles. **(b)** Flow curves from (a), rescaled according to Eq. (6).

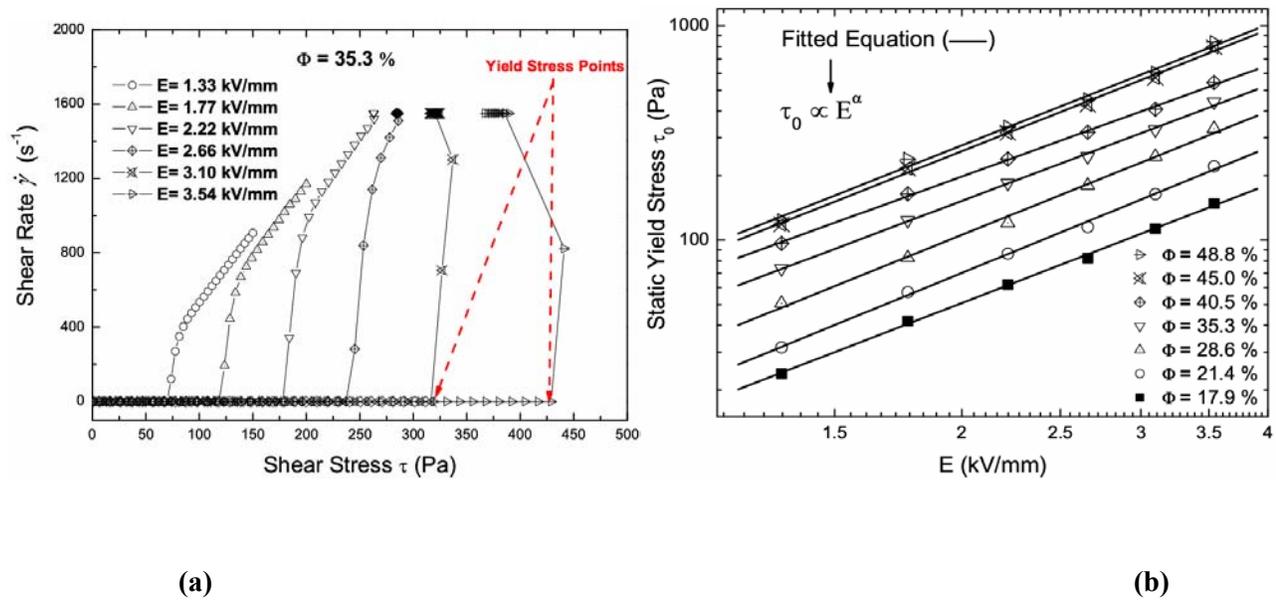

**(a)** **(b)**

**Fig. 6: (a)** Curves of shear rate vs. shear stress recorded to determine the static yield stress of a suspensions of concentration Φ=35% under various electric field strengths. **(b)** Log-log plot of all the static yield stress values as a function of the applied electric field, for different concentrations of the suspensions.

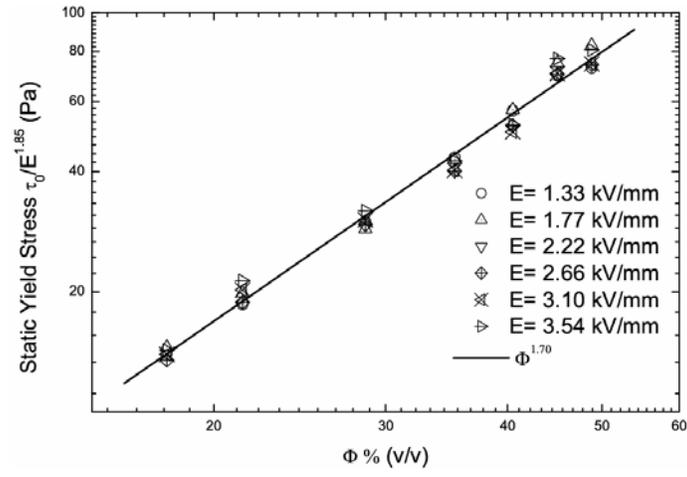

**Fig. 7:** Log-log plot of the static yield stress, normalized by $E^{1.86}$, vs. the volume fraction at different strengths of the applied electric field. The power law fitted to the whole data, in the form $\Phi^{1.70}$, is also plotted.

| ER Fluids → | Our Sample | Mica [16] | Hematite[41] | Saponite[42] | Zeolite[43] | GER[44] |
|---|---|---|---|---|---|---|
| Φ → | 17.9% (v/v) | 15%(v/v) | 15%(v/v) | 0.11g/ml | 30%(v/v) | 30%(v/v) |
| $\tau_0$ (Pa) → | ~20 | ~100 | ~85 | ~50 | ~3000 | ~15000 |

**Table 1**: A comparison of static yield stress values for various ER fluids including that addressed in the present paper, under an applied electric field of about 1.0 kV/mm.

| Φ % (v/v) | 17.9 | 21.4 | 28.6 | 35.3 | 40.5 | 45.0 | 48.8 |
|---|---|---|---|---|---|---|---|
| α | 1.83±0.03 | 1.94±0.04 | 1.91±0.06 | 1.79±0.03 | 1.72±0.03 | 1.88±0.05 | 1.87±0.06 |

**Table 2**: Values obtained for the parameter α in Fig. 6(b), at various volume fractions Φ. The strength of the applied electric field was between 1.33 and 3.54 kV/mm.